\documentclass[aps,pra,reprint,superscriptaddress]{revtex4-1}

\usepackage{amsmath}
\usepackage{amsfonts}
\usepackage{braket}
\usepackage{graphicx}
\usepackage{hyperref}
\usepackage{color}

\usepackage[caption=false]{subfig}

\allowdisplaybreaks[1]

\DeclareMathOperator{\per}{\textrm{Per}}
\DeclareMathOperator{\symm}{\textrm{S}}
\DeclareMathOperator{\unitary}{\textrm{U}}
\DeclareMathOperator{\tdist}{\delta_\textrm{tr}}
\DeclareMathOperator{\trace}{\textrm{Tr}}
\DeclareMathOperator{\var}{\textrm{Var}}
\DeclareMathOperator{\cov}{\textrm{Cov}}
\DeclareMathOperator{\poly}{\textrm{Poly}}
\DeclareMathOperator{\prob}{\textrm{Pr}}
\DeclareMathOperator{\mean}{\mathbb{E}}

\DeclareGraphicsExtensions{.pdf}

\begin{document}
\title{Classically simulating near-term partially-distinguishable and lossy boson sampling} 
\author{Alexandra E. Moylett}
\email{alex.moylett@bristol.ac.uk}
\affiliation{Quantum Engineering Technology Labs, H. H. Wills Physics Laboratory and Department of Electrical \& Electronic Engineering, University of Bristol, BS8 1FD, UK}
\affiliation{Quantum Engineering Centre for Doctoral Training, H. H. Wills Physics Laboratory and Department of Electrical \& Electronic Engineering, University of Bristol, BS8 1FD, UK}
\affiliation{Heilbronn Institute for Mathematical Research, University of Bristol, BS8 1SN, UK}
\author{Ra\'{u}l Garc\'{i}a-Patr\'{o}n}
\email{rgarciap@ulb.ac.be}
\affiliation{Centre for Quantum Information and Communication, Ecole Polytechnique de Bruxelles, CP 165, Universit\'{e} Libre de Bruxelles, 1050 Brussels, Belgium}
\author{Jelmer J. Renema}
\email{j.j.renema@utwente.nl}
\affiliation{Complex Photonic Systems (COPS), Mesa+ Institute for Nanotechnology, University of Twente, PO Box 217, 7500 AE Enschede, The Netherlands}
\author{Peter S. Turner}
\email{peter.turner@bristol.ac.uk}
\affiliation{Quantum Engineering Technology Labs, H. H. Wills Physics Laboratory and Department of Electrical \& Electronic Engineering, University of Bristol, BS8 1FD, UK}

\date{\today}

\begin{abstract}
Boson Sampling is the problem of sampling from the same distribution as indistinguishable single photons at the output of a linear optical interferometer. 
It is an example of a non-universal quantum computation which is believed to be feasible in the near term and cannot be simulated on a classical machine.
Like all purported demonstrations of ``quantum supremacy'', this motivates optimizing classical simulation schemes for a realistic model of the problem, in this case Boson Sampling when the implementations experience lost or distinguishable photons.
Although current simulation schemes for sufficiently imperfect boson sampling are classically efficient, in principle the polynomial runtime can be infeasibly large.
In this work, we develop a scheme for classical simulation of Boson Sampling under uniform distinguishability and loss, based on the idea of sampling from distributions where at most $k$ photons are indistinguishable.
We show that asymptotically this scheme can provide a polynomial improvement in the runtime compared to classically simulating idealised Boson Sampling. 
More significantly, we show that in the regime considered experimentally relevant, our approach gives an substantial improvement in runtime over other classical simulation approaches.
\end{abstract}

\maketitle

\section{Introduction}
\label{sec:intro}

Since the idea of quantum computers was first proposed, speedups have been theoretically demonstrated for a variety of problems~\cite{montanaro2016}. 
More recently, this has turned to interest in near term quantum speedups on realistic hardware~\cite{harrow2017}. 
Boson Sampling is an example of such a problem~\cite{aaronson2011}, which considers the complexity of sampling from the same probability distribution as $n$ collision-free indistinguishable photons output from an $m$-mode linear optical interferometer drawn uniformly at random.
Aaronson and Arkhipov showed that the ability to exactly simulate Boson Sampling in polynomial time would imply that $P^{\#P} = BPP^{NP}$, and the Polynomial Hierarchy would collapse to the third level. 
This is because the output probabilities of a linear optical interferometer are proportional to the permanent of an $n \times n$ complex matrix \cite{scheel2008}, which is $\#P$-Hard to compute.
The same result was also shown for approximately sampling from the same distribution, modulo two conjectures related to the permanents of Gaussian matrices~\cite{aaronson2011}. 
This led to significant interest in Boson Sampling, with experimental demonstrations ranging from two to six photons~\cite{broome2013, spring2013, tillmann2013, crespi2013, carolan2015, wang2017, zhong2018, paesani2018}. 

This has driven interest in improving classical simulations of Boson Sampling, in order to verify any speedup gained from an experimental implementation.
Neville \textit{et al.}~\cite{neville2017} showed that Boson Sampling can be classically performed for 30 bosons across 900 modes at a rate of 250 samples in under five hours on a standard laptop. 
This was further improved by Clifford and Clifford~\cite{clifford2017}, who showed that sampling from $n$ photons across $m$ modes can be classically performed in $O(n2^n + \poly(n,m))$ time and $O(m)$ space.

Recently classical simulations have been expanded to consider practical issues such as photon distinguishability, based on a rich collection of theoretical work in this area \cite{deguise2014, tamma2014, shchesnovich2014, rohde2015, shchesnovich2015, tamma2015, laibacher2015, tichy2015, tillmann2015, menssen2017, laibacher2017, spagnolo2014, giordani2018}. 
Renema \textit{et al.}~\cite{renema2018} demonstrated that Boson Sampling with partially-distinguishable photons can be simulated in time which grows polynomially with $n$. 
This was later expanded to consider loss as well~\cite{renema2018loss}. 
However, the runtime might still not be efficient in practice, as the polynomial can be large. There is also a further disadvantage in that the error bounds are the average case for a random linear optical interferometer, meaning that there could be interferometers for which the algorithm performs significantly worse.
A significant improvement could be achieved through adapting the method of Clifford \& Clifford to this algorithm, but there are challenges with this approach.

Another motivation for adapting Clifford \& Clifford to photonic imperfection is the task of classically simulating other photonic regimes, such as Gaussian Boson Sampling \cite{hamilton2017}. The probability distribution of Gaussian Boson Sampling is that of $n$ indistinguishable squeezed states at the output of an $m$-mode linear optical interferometer, and depends on the Hafnian of a matrix. Unlike Boson Sampling, there is no known polynomial time classical algorithm for computing the outcome of $n$ fully distinguishable squeezed states. On the other hand, it is classically efficient to sample distinguishable squeezed states, via a similar approach to that used for classically sampling distinguishable single photons \cite{aaronson2014}. As a result, adapting the Clifford \& Clifford algorithm to non-ideal Boson Sampling models provides a first step towards being able to classically simulate imperfections in Gaussian Boson Sampling.

Here we consider the cost of classically simulating Boson Sampling when the photons are partially distinguishable or lossy.
We look at the same model of distinguishability as considered in \cite{renema2018,renema2018loss}, and use techniques for modelling photon distinguishability in first quantization \cite{moylett2018, stanisic2018} to show that this is akin to choosing the indistinguishable photons of a Boson Sampling experiment via the binomial distribution. 
We combine this with the well-studied model of uniform loss, which also follows a binomial distribution. 
This gives rise to a method which is able to naturally apply the Clifford \& Clifford algrithm and take advantage of its efficiency. This algorithm also offers a worst-case error bound for \textit{any} linear optical interferometer, rather than a random one.
Although this approach only offers a polynomial improvement compared to the runtime for ideal Boson Sampling (unlike the exponential improvement shown in \cite{renema2018,renema2018loss}) we use analytical bounds to show that for photon numbers of experimental interest our algorithm can make a significant improvement over alternative approaches. 

This article is laid out as follows.
In Sec.\ \ref{sec:background}, we give an overview of distinguishability in first quantization, and summarise the work of classical simulation algorithms including \cite{renema2018, renema2018loss}. 
In Sec.\ \ref{sec:expansion}, we show what the model of distinguishability looks like in first quantization, and provide an alternative classical simulation. 
In Sec.\ \ref{sec:average-case}, we consider average error bounds for a Haar-random unitary interferometer, via the methods explained in Sec.\ \ref{sec:renema-review}. 
In Sec.\ \ref{sec:worst-case}, we improve this bound to a worst-case error bound, by computing an upper bound for the trace distance between our approximation and the model. 
In Sec.\ \ref{sec:loss}, we expand these results to consider uniform loss, and show how distinguishability and loss relate to each other. 
In Sec.\ \ref{sec:empirical-errors}, we explore these error bounds for experimentally interesting numbers of photons, and show that there are some cases where our algorithm offers an improvement. 
Finally, we briefly consider \emph{non}-uniform loss, where loss is is a function of the number of optical components, and use the methods of \cite{garciapatron2017, oszmaniec2018} to show that classical simulations with non-uniform loss also become easier when distinguishability is introduced.
We conclude with some open research questions in Sec.\ \ref{sec:conclusion}.

\section{Preliminaries}
\label{sec:background}

\subsection{Partial distinguishability in first quantization}

Studying photonics in first quantization allows an arbitrary linear optical experiment to be expressed in the quantum circuit model ~\cite{moylett2018, stanisic2018, oszmaniec2018}.
Bosonic symmetrisation can be implemented efficiently, e.g. through use of the quantum Schur transform \cite{bacon2007}, and the main sources of errors in linear optics can be modelled: loss as erasure of qudits and distinguishability as correlation to extra photonic degrees of freedom, both of which lead to decoherence of the quantum circuit.

We start by considering arbitrary distinguishability of photonic scattering, models of which have been proposed in \cite{tamma2015, laibacher2015, tichy2015}. 
Let $m,n$ be integers such that $m\in O(n^2)$. 
Let $U\in \unitary(m)$ be an $m$-mode linear optical interferometer, and let $S,S'$ be $n$-photon $m$-mode Fock vectors, meaning that $S=(S_0,\dots,S_m)$, $S'=(S'_0,\dots,S'_m)$, and $\sum_{i=0}^mS_i=\sum_{i=0}^mS'_i=n$. 
The probability of seeing output occupation $S'$ from input $S$ with interferometer $U$ is
\begin{equation}\label{eq:TichyDist}
\prob[S'] =  \frac{1}{\prod_{i=1}^m S_i! S'_i!} \sum_{\tau, \tau' \in \textrm{S}_n} \prod_{k=1}^n U_{s'_k,s_{\tau(k)}} U^\ast_{s'_k,s_{{\tau'}(k)}} \mathcal{S}_{{\tau'(k)},{\tau(k)}},
\end{equation}
where $\mathcal{S}_{k,l}$ is an $n \times n$ Gram matrix describing the (in)distinguishability of pairs of photons, $S$ is a Fock array describing the arrangement of input photons, $s$ is the corresponding array of single particle states in first quantisation, likewise for $S'$ and $s'$, $U$ is a $m\times m$ unitary matrix describing the interferometer, and S$_n$ is the group of permutations of $n$ particles.
Of particular note is that when $\mathcal{S}$ is the all-1 matrix, we have
\begin{align}
\prob[S'] &= \frac{1}{\prod_{i=1}^m S_i! S'_i!} \sum_{\tau, \tau' \in \textrm{S}_n} \prod_{k=1}^n U_{s'_k,s_{\tau(k)}} U^\ast_{s'_k,s_{{\tau'}(k)}}\\
&= \frac{1}{\prod_{i=1}^m S_i! S'_i!} \sum_{\tau \in \textrm{S}_n} \prod_{k=1}^n U_{s'_k,s_{\tau(k)}} \sum_{\tau' \in \textrm{S}_n} \prod_{k=1}^nU^\ast_{s'_k,s_{{\tau'}(k)}}\\
&= \frac{1}{\prod_{i=1}^m S_i! S'_i!} \per(U_{S,S'})\per(U^\ast_{S,S'})\\
&=  \frac{1}{\prod_{i=1}^m S_i! S'_i!} |\per(U_{S,S'})|^2,
\end{align}
where $U_{S,S'}$ is a matrix defined by taking rows and columns of $U$ according to $S$ and $S'$~\cite{scheel2008}. 
This is the output probability distribution for Boson Sampling with indistinguishable photons. 
Likewise, we can see that when $\mathcal{S}$ is the identity matrix, we have
\begin{align}
\prob[S'] &= \frac{1}{\prod_{i=1}^m S_i! S'_i!} \sum_{\tau \in \textrm{S}_n} \prod_{k=1}^n U_{s'_k,s_{\tau(k)}} U^\ast_{s'_k,s_{{\tau}(k)}}\\
&= \frac{1}{\prod_{i=1}^m S_i! S'_i!} \sum_{\tau \in \textrm{S}_n} \prod_{k=1}^n |U_{s'_k,s_{\tau(k)}}|^2\\
&=  \frac{1}{\prod_{i=1}^m S_i! S'_i!} \per(|U_{S,S'}|^2),
\end{align}
which is the distribution for fully distinguishable photons. 
In first quantisation, the input state can be written as~\cite{moylett2018}
\begin{equation}
\rho
=\frac{1}{n!\prod_{i=1}^m S_i!}\sum_{\sigma,\sigma'\in\textrm{S}_n}\sigma|s\rangle\langle s|\sigma'^\dagger \prod_{k=1}^n \mathcal{S}_{\sigma'^{-1}(k),\sigma^{-1}(k)}, \label{eqn:dist-state}
\end{equation}
For convenience, we will assume here and throughout the rest of this paper that our photons start in the Fock state $\ket{S}=\ket{1^n0^{m-n}}$. 
We will use the corresponding first quantized states of this form in the following.

\subsection{Classical simulation of fully (in)-distinguishable lossless Boson Sampling}

Boson Sampling under ideal conditions (lossless indistinguishable single photons) is intractable for sufficiently large $n$. 
Until recently the only classical simulation method explicitly known was to compute the entire probability distribution before taking a sample, though it was widely believed that more efficient, albeit still exponential time, approaches existed. 
A brute force method cannot scale, due to both the number of possible outcomes and the complexity of computing even one $n\times n$ complex matrix permanent.

Two major results gave the first explicit classical simulation strategies which were faster than brute-force sampling. 
The first, by Neville \textit{et al.}~\cite{neville2017}, demonstrated that Boson Sampling experiments with up to 30 photons could be simulated on a single laptop, and suggests that a supercomputer could handle up to 50 photons. 
This was achieved by starting with the classical distribution of $n$ distinguishable photons, and then using Metropolised Independence Sampling to adapt the distribution to that of ideal Boson Sampling.
The second result, by Clifford \& Clifford \cite{clifford2017}, gave a classical algorithm for exact Boson Sampling and runs in time equivalent to computing two $n \times n$ matrix  permanents per sample with a polynomial overhead. 
This is through a combination of optimizations, particularly computing marginal probabilities and sampling via the chain rule.
Our approach here is to make these more efficient techniques applicable to realistic situations with distinguishability and loss.

It is also worth discussing at this point how to classically simulate boson sampling with completely distinguishable photons, for which a polynomial time exists \cite{aaronson2014}. In this case, there is no photon interference, so photons can be sampled individually. This is done by taking a photon which starts in mode $i$, and sampling output mode $j$ with probability $|U_{j,i}|^2$. Repeating for all photons gives us the complete sample in $O(mn)$ time.

\subsection{Expanding in terms of fixed points}
\label{sec:renema-review}

In \cite{renema2018, renema2018loss}, Renema \textit{et al.}\ consider a model where inter-photon distinguishability is measured by an inner product of pure states \cite{tamma2014, deguise2014, shchesnovich2015, rohde2015, tamma2015, tichy2015}. 
The probability distribution of arbitrarily distinguishable bosons is modelled as
\begin{equation}
\prob[S'] = \sum_{\sigma\in\symm_n}\prod_{i=1}^n\mathcal{S}_{i,\sigma(i)}\per(M*M^*_{1,\sigma}) ,
\end{equation}
where $\mathcal{S}$ is the same matrix describing the distinguishability as in the previous section, $M$ is a matrix defined by the rows and columns of our interferometer $U$ selected based on our photon output $S'$ and input $S$, $M^*_{1,\sigma}$ is the conjugate matrix with the identity permutation applied to rows and permutation $\sigma$ applied to columns, and $*$ denotes element-wise multiplication. 
They further restrict to a model where the indistinguishability overlap is defined by a single parameter $\mathcal{S}_{i,j} = x + (1-x)\delta_{i,j}, x \in [0,1]$.
The sum over permutations can be ordered based on how many \emph{fixed points} a permutation has, giving
\begin{equation}
\prob[S'] = \sum_{j=0}^n\sum_{\sigma^j}x^j\per(M*M^*_{1,\sigma}) , \label{eqn:renema-state}
\end{equation}
where $\sigma^j$ denotes permutations which have $n-j$ elements as fixed points.
Each permanent can be broken down via the Laplace expansion into a sum of a complex matrix permanent multiplied by a positive matrix permanent:
\begin{equation}
\prob[S'] = \sum_{j=0}^n\sum_{\sigma^j}x^j\sum_{\substack{J'\leq S'\\|J'|=j}}\per(M_{J',1}*M^*_{J',\sigma_p})\per(|M_{\bar{J'},\sigma_u}|^2) .
\end{equation}
Here we are now choosing submatrices of $M$, with $J'$ representing the $\binom{n}{j}$ possible combinations of rows from $M$, $\bar{J'}$ representing the remaining rows, and $\sigma_p$ and $\sigma_u$  representing permuted and unpermuted elements of $\sigma$ respectively.
The $J' \leq S'$ notation is used to indicate that $J'$ is a Fock state such that $J_i\leq S'_i \,\forall i\in[m]$.

The classical simulation method used truncates the number of fixed points in a permutation as being at most $k$, with the remainder of the probability treated as an error margin. 
It is important to note while these approximations are real, they are not necessarily positive. 
This is due to the truncation, where positive higher order terms which would have corrected the probability to be positive are now missing from the approximation. 
To correct this, any negative approximations are rounded up to $0$.
These probabilities are then used to train a Metropolised Independence Sampler, akin to the technique of \cite{neville2017}. 
Training this sampler requires approximating a number of probabilities dependent on the underlying distribution, each of which involves computing $O(n^{2k})$ permanents of $k\times k$ complex matrices, and the same number of permanents of $(n-k)\times (n-k)$ matrices with non-negative entries. 
The permanents of $k\times k$ complex matrices can be computed classically in $O(k2^k)$ time via Ryser's algorithm, and the permanents of matrices with non-negative entries can be approximated up to multiplicative error in polynomial time \cite{jerrum2004,huber2008}. 
As long as $k$ is independent of $n$, this means that there is a polynomial runtime.

To work out a suitable value $k$, define coefficients $c_j$ as
\begin{equation}
c_j = \sum_{\substack{J'\leq S'\\|J'|=j}}\per(M_{J',1}*M^*_{J',\sigma_p})\per(|M_{\bar{J'},\sigma_u}|^2)\label{eqn:coefficients}.
\end{equation}
Assuming the matrices are Gaussian, the variance of each permanent can be bounded as

\begin{equation}
\var[\per(M_{J',1}*M^*_{J',\sigma_p})] = \frac{j!}{m^{2j}},
\end{equation}
and
\begin{equation}
\var[\per(|M_{\bar{J'},\sigma_u}|^2)] < \frac{(n-j)!}{m^{2(n-j)}}\sum_{l=0}^{n-j}\frac{1}{l!}.
\end{equation}

This leads to two key results. The first is that the variance of $c_j$ tends towards a constant value:
\begin{align}
\var[c_j] &< \left(\frac{n!}{m^n}\right)^2\frac{1}{e}\sum_{l=0}^{n-j}\frac{1}{l!} \label{eqn:var}\\
&\rightarrow\left(\frac{n!}{m^n}\right)^2 \textrm{ as } n\rightarrow\infty,
\end{align}
and the second is that the covariance for different values of $j$ is zero:
\begin{equation}
\cov[c_j,c_j'] = 0\, \forall\, j\neq j'\label{eqn:covar}.
\end{equation}
From this one can approximate the variance of the error as a geometric series, which as $n\rightarrow\infty$ tends towards the inequality
\begin{align}
\var[\Delta \prob[S']] &= \var[\prob[S']-\prob_k[S']]\\
&= \var\left[\sum_{j=k+1}^nx^j\var[c_j]\right]\label{eqn:renema-variance}\\
&< \left(\frac{n!}{m^n}\right)^2\left(\frac{x^{2(k+1)}}{1-x^2}\right),
\end{align}
\noindent where $P_k$ is the probability distribution when truncated at $j\leq k$.

Finally one can use a Markov inequality to show that if the variance of the error is of the form $(n!/m^n)^2\epsilon^2$, the average error of the simulation is at most $\epsilon$ \cite{renema2018loss}.
Crucially, this value of $\epsilon$ is only dependent on $x$ and $k$ and no longer dependent on $n$.
This means that for any value of $x$, one can choose a suitable value of $k$ to achieve a required error $\epsilon$, and run a classical simulation in time polynomial in $n$.

Although this runtime is polynomial in terms of $n$ and can therefore be considered asymptotically efficient, it might not be classically simulable in practice. 
There are three main contributions to this: 
First, the algorithm is reliant on Metropolised Independence Sampling, which potentially requires many probabilities to be approximated per sample. 
Second, approximating each probability requires $O(n^{2k})$ permanents of $k\times k$ matrices, which even for small $k$ could be a large number of permanents. 
And third, approximating each probability requires $O(n^{2k})$ permanents of $(n-k)\times(n-k)$ matrices with non-negative elements. 
Although approximating permanents of matrices with non-negative elements can be achieved in polynomial time, classical algorithms still have a runtime ranging from $O((n-k)^4\log(n-k))$ to $O((n-k)^{7}\log^4(n-k))$, depending on the sparsity of the matrix~\cite{huber2008}. These issues are the main points to address in order to achieve a practical classical algorithm for Boson Sampling. Clifford \& Clifford could help to alleviate these issues, but there is a challenge due to the fact that the approximation used in Renema et al.\ does not correspond to a bosonic state. 
This in turn leads to negative probabilities, which are not clear how to correct for the Clifford \& Clifford algorithm.

\subsection{Classical simulation of lossy Boson Sampling}

Another common source of imperfections in linear optics is that of photon loss, which arises through a number of different means. 
Indeed, any large-scale demonstration of Boson Sampling is bound to face photon loss, and therefore needs to take such issues into account. 
Some results have already shown instances where hardness is still retained, such as when only a constant number of photons are lost \cite{aaronson2016,wang2018}.

Neville \textit{et al.}\ compared the classical simulation of their approach to a Boson Sampling experiment where any photon loss was considered a rejected experiment \cite{neville2017}. 
In \cite{renema2018loss}, the method described in Sec.\ \ref{sec:renema-review} was adapted to consider uniform loss, showing that the same result can be found, with the only difference being that $x$ is now replaced by $\alpha=\sqrt{\eta}x$, where $\eta$ is the probability of each individual photon surviving. 
Crucially, this result demonstrated that Boson Sampling where a constant fraction of photons were lost can be simulated in $O(\ell^{2k}k2^k)$, where $\ell$ is the number of photons which survive and $k$ is only dependent on the constant $\ell/n$, distinguishability $x$, and the desired accuracy of the simulation. 
This can be expanded to classically simulating Boson Sampling under uniform loss by sampling $\ell$ from the binomial distribution before sampling output photons, which offers a runtime of $O(n^{2k}k2^k)$. 
Novel classical simulations for Boson Sampling under loss have also been considered by use of classically simulable states such as thermal \cite{garciapatron2017} or separable \cite{oszmaniec2018} states.

There has also been some consideration of how classical simulations can be generalised to non-uniform loss. 
This usually means photon loss that is dependent on the number of optical components, with each component having transmission probability $\tau$. 
Classical simulation methods can be generalised to this model by identifying a layer of uniform losses from the circuit, followed a non-uniform lossy circuit which can be simulated classically through the use of additional modes for lost photons \cite{garciapatron2017,oszmaniec2018}. 
These results showed that Boson Sampling under non-uniform loss can be classically simulated as long as the smallest number of components a photon encounters is logarithmic in $n$.

\section{Expanding in terms of states}
\label{sec:expansion}

We will now introduce a new expression for partially distinguishable particles. We start by writing the input state of $n$ photons, with pairwise distinguishability parameter $x$ as in the previous section, in first quantization
\begin{equation}
\label{eqn:rhon}
\rho_{n,x} = \frac{1}{n!}\left(\sum_{\sigma,\sigma'\in\symm_n}\sigma\ket{s}\bra{s}\sigma'x^{\sigma\cdot\sigma'}\right),
\end{equation}
where we have used $\sigma\cdot\sigma'$ to denote the number of places where permutations $\sigma$ and $\sigma'$ match.
For reference, the expansion of \cite{renema2018,renema2018loss} is carried out by identifying $\sigma$ and $\sigma'$ that match for a fixed set of $i$ points:
\begin{equation}
\label{eqn:rhoni}
\rho_{n,x} = \frac{1}{n!}\left(\sum_{i=0}^nx^i\sum_{\substack{\sigma,\sigma'\in\symm_n\\\exists I\subseteq [n] \textrm{ s.t. } |I|=i\\\sigma^{-1}(j)\neq\sigma'^{-1}(j)\forall j \in I\\\sigma^{-1}(j)=\sigma'^{-1}(j)\forall j \notin I}}\sigma\ket{s}\bra{s}\sigma'^\dagger\right).
\end{equation}
Note here that the sums over permutations do not correspond to physical states. 
This can be seen by the fact that for $i\neq 0$ this summation has no elements along the diagonal of the density matrix, as $\sigma$ and $\sigma'$ need to differ in \emph{exactly} $i$ places.

We instead look at an alternative expansion, in order to decompose the model into a linear combination of physical states:
\begin{align}
\rho_{n,x} &= \frac{1}{n!}\left(\sum_{i=0}^np_i\sum_{\substack{\sigma,\sigma'\in\symm_n\\\exists I\subseteq [n] \textrm{ s.t. } |I|=i\\\sigma^{-1}(j)=\sigma'^{-1}(j)\forall j \notin I}}\sigma\ket{s}\bra{s}\sigma'^\dagger\right)\label{eqn:state-expansion}\\
&= \sum_{i=0}^np_i\sum_{\substack{I\subseteq[n]\\|I|=i}}\left(\frac{1}{n!}\sum_{\substack{\sigma,\sigma'\in\symm_n\\\sigma^{-1}(j)=\sigma'^{-1}(j)\forall j \notin I}}\sigma\ket{s}\bra{s}\sigma'^\dagger\right)\\
&= \sum_{i=0}^np_i\sum_{\substack{I\subseteq[n]\\|I|=i}}\rho_I, \label{eq:newrho}
\end{align}
where $\rho_I$ is the state where photons in modes $j \in I$ are fully indistinguishable from each other, all other photons are fully distinguishable, and $0 \leq p_i \leq 1$ is a coefficient dependent on $x$ and $n$ determining the probability of a state with $i$ indistinguishable single photons.

Note that unlike Eq.\ (\ref{eqn:rhoni}), where permutations must differ in exactly $i$ points, in Eq.\ (\ref{eqn:state-expansion}) we allow permutations to differ in \emph{at most} $i$ points. 
This means that elements closer to and along the diagonal of the density matrix are also part of this summation, and this means that each sum over $\sigma,\sigma' \in \symm_n$ forms a valid density matrix.

Already we can see how a classical simulation might work -- if we are able to sample $p_i$ efficiently, then we can choose $\rho_I$ by selecting $i$ photons uniformly at random to be indistinguishable. 
These $i$ photons can be classically simulated using Clifford \& Clifford \cite{clifford2017}, while the remaining $n-i$ photons are treated as fully distinguishable photons, each of which can be simulated individually in polynomial time \cite{aaronson2014,neville2017}.

\subsection{The $p_i$ are binomially distributed}
\label{sec:understanding-pi}

Here, we will show that the coefficients $p_i$ follow the binomial distribution
\begin{equation}
p_i = x^i(1-x)^{n-i}.
\end{equation}


To see that the matrix elements of Eq.~(\ref{eq:newrho}) with $p_i$ binomially distributed equal those of Eq.~(\ref{eqn:rhoni}), consider  $\sigma,\sigma'$ which differ at points in the set $I$, where $|I|=i$; the coefficient here should be $x^i$.
Contributing to this element of the density matrix will be the state $\rho_I$, as well as other states $\rho_{I'}$, where $I \subseteq I'$. 
The number of such sets $I'$ is $\binom{n-i}{i'-i}$, as it is equivalent to choosing $i'$ from $n$ elements when $i$ elements have already been chosen. 
The corresponding matrix element is
\begin{align}
&\frac{1}{n!}\left(\sum_{{i'}=i}^n x^{i'}(1-x)^{n-{i'}}\binom{n-i}{{i'}-i}\right) \\
&= \frac{1}{n!}\left(x^i\sum_{{i'}=i}^n x^{{i'}-i}(1-x)^{n-{i'}}\binom{n-i}{{i'}-i}\right)\\
&= \frac{1}{n!}\left( x^i\sum_{{i'}=0}^{n-i}x^{i'}(1-x)^{n-i-{i'}}\binom{n-i}{{i'}}\right)\\
&= \frac{x^i(x + 1 - x)^{n-i}}{n!}\\
&= \frac{x^i}{n!}.
\end{align}
It is not hard to see that the state is normalised
\begin{align}
\trace[\rho_{n,x}] &= \sum_{i=0}^nx^i(1-x)^{n-i}\sum_{\substack{I\subseteq[n]\\|I|=i}}\trace[\rho_I]\\
&= \sum_{i=0}^nx^i(1-x)^{n-i}\sum_{\substack{I\subseteq[n]\\|I|=i}}1\\
&= \sum_{i=0}^nx^i(1-x)^{n-i}\binom{n}{i}\\
&= (x + 1 - x)^n\\
&= 1.
\end{align}
Thus this model of fixed pairwise distinguishability can be written as an expansion in terms of valid states, where indistinguishable photons are drawn from a binomial distribution.

\subsection{Classical simulation} 

We can now see explicitly how a simulation for Boson Sampling with distinguishable photons would work. 
First, we sample an integer $i \in [n]$ according to the Binomial distribution with coefficients $n$ and $x$. 
Next, we sample a subset $I$ of the photons uniformly at random from the $\binom{n}{i}$ possible subsets of size $i$. 
These are the indistinguishable photons of our simulation, which we simulate using Clifford and Clifford in $O(i2^i + \poly(i,m))$ time. 
The remaining $n-i$ photons are considered to be distinguishable. 
Rather than needing to compute the output probabilities of these photons colletively, which could take between $O(n-i)^4\log(n-i)$ and $O(n-i)^7\log^4(n-i)$ time via permanents of matrices with non-negative entries \cite{huber2008}, we can instead sample each distinguishable photon individually. 
To do so, we take a distinguishable photon in mode $a$, and compute the probability of this photon being measured in mode $b$ as $|U_{b,a}|^2$. 
Thus we can compute all output probabilities and obtain a sample for a single distinguishable photon in $O(m)$ time, meaning that we can obtain a sample for all $n-i$ distinguishable photons in $O(m(n-i))$ time \cite{aaronson2014,neville2017}.

The run time is dominated by the time taken to sample our indistinguishable photons, which can be as large as $O(n2^n + \poly(n,m))$ if we are unlucky. 
By truncating our Binomial sampling up to some level $k$, we can simulate Boson Sampling up to some level of error. 
The extent of this error will be the focus of Secs. \ref{sec:average-case} \& \ref{sec:worst-case}.

\section{Average case error}
\label{sec:average-case}

We can use the same strategies used in \cite{renema2018,renema2018loss} to derive an error bound for Boson Sampling via state truncation for a Haar-random interferometer. We shall do this by considering the total variation distance between our approximation and the model for partial distinguishability for a Gaussian matrix. This is given by

\begin{equation}
\mean[\Delta P] = \sum_{S'}\mean|\prob[S']-\prob_k[S']|,
\end{equation}
where $\prob_k$ is the probability distribution truncated at $k$ indistinguishable photons via our approximation. For a specific outcome $S'$, we can expand the right hand side to
\begin{align}
\mean\left[|\prob[S']-\prob_k[S']|\right] = &\mean\left|\sum_{i=0}^k\left(p_i-p_i'\right)\sum_{\substack{I\subseteq[n]\\|I|=i}}P_I[S']\right.\nonumber\\
&\quad+\left.\sum_{i=k+1}^np_i\sum_{\substack{I\subseteq[n]\\|I|=i}}P_I[S']\right|,
\end{align}
where $P_I$ is now the probability distribution with indistinguishable photons defined by set $I$, and $p_i'=p_i/(\sum_{i=0}^kx^i(1-x)^{n-i}\binom{n}{i})$ is the normalised version of the $p_i$ coefficients defined in Sec.\ \ref{sec:expansion}. 
Note that $P_I$ is akin to the distribution arising from state $\rho_I$ in Sec.\ \ref{sec:expansion}. 
We can use the triangle inequality to bound this value to
\begin{align}
\mean\left[|\prob[S']-\prob_k[S']|\right] &\leq \mean\left|\sum_{i=0}^k\left(p_i-p_i'\right)\sum_{\substack{I\subseteq[n]\\|I|=i}}P_I[S']\right|\nonumber\\
&+\mean\left|\sum_{i=k+1}^np_i\sum_{\substack{I\subseteq[n]\\|I|=i}}P_I[S']\right|\\
&= \mean[\Delta P_{\leq k}] +\mean[\Delta P_{> k}],
\end{align}
where we have introduced variables $\Delta P_{\leq k}$ and $\Delta P_{> k}$ for convenience. 
We shall consider the expected values of these terms for a Gaussian matrix separately, starting with the latter.
Using the Laplace expansion, we find that
\begin{align}
\mean[\Delta P_{>k}] &= \sum_{i=k+1}^np_i\sum_{\substack{I\subseteq[n]\\|I|=i}}\mean\left[P_I[S']\right],\\
&=\sum_{i=k+1}^np_i\sum_{\substack{I\subseteq[n]\\|I|=i}}\sum_{\substack{J\leq S'\\|J|=i}}\mean\left[|\per(U_{I,J})|^2\per(|U_{\bar{I},\bar{J}}|^2)\right],\label{eqn:laplace}
\end{align}
where $U_{I,J}$ is a matrix defined from our interferometer $U$ by selecting columns according to $I$ and rows according to $J$, and $|U_{\bar{I}\bar{J}}|^2$ is a matrix whose elements are the absolute values squared of $U_{\bar{I},\bar{J}}$.

We next need to consider the expected values of the matrix permanents in Eq.\ (\ref{eqn:laplace}) for a Haar random unitary. 
To do this, we shall assume the matrix describing our interferometer is Gaussian. 
This allows us to assume that each entry of $U_{I,J}$ and $|U_{\bar{I},\bar{J}}|^2$ is independent, and that the two matrices are independent of each other. 
Starting with $\per(|U_{\bar{I},\bar{J}}|^2)$, we note that this is a Gaussian matrix of size $(n-i)\times(n-i)$, and each entry is the square of two independent Gaussians, meaning that each entry of $|U_{\bar{I},\bar{J}}|^2$ has expected value $1/m$. 
From this, we can calculate the expected value as
\begin{equation}
\mean[\per(|U_{\bar{I},\bar{J}}|^2)] = \frac{(n-i)!}{m^{n-i}}.
\end{equation}

For $U_{I,J}$, we note that each element of $U_{I,J}$ is an independent Gaussian entry, with mean value $0$ due to symmetry, and second order moment $\mean[|U_{i,j}|^2] = 1/m$. The second order moment for the permanent can then be calculated using the same methods as in \cite{aaronson2011}:
\begin{align}
\mean[|\per(U_{I,J})|^2] &= \mean\left[\sum_{\sigma,\sigma'\in\symm_i}\prod_{l=1}^n(U_{I,J})_{l,\sigma(l)}(U^*_{I,J})_{l,\sigma'(l)}\right]\\
&= \mean\left[\sum_{\sigma\in\symm_i}\prod_{l=1}^n|(U_{I,J})_{l,\sigma(l)}|^2\right]\\
&= \sum_{\sigma\in\symm_i}\prod_{l=1}^n\mean\left[|U_{l,\sigma(l)}|^2\right]\\
&= \frac{i!}{m^{i}}.
\end{align}
Because the two matrices are independent~\cite{renema2018}, we can express the expected value of their product as
\begin{align}
\mean\left[|\per(U_{I,J})|^2\per(|U_{\bar{I},\bar{J}}|^2)\right] &= \frac{(n-i)!}{m^{n-i}}\times\frac{i!}{m^{i}}\\
&= \frac{(n-i)!i!}{m^n}.
\end{align}
Plugging this into Eq.\ (\ref{eqn:laplace}), we find that
\begin{align}
\mean[\Delta P_{>k}] &= \sum_{i=k+1}^np_i\sum_{\substack{I\subseteq[n]\\|I|=i}}\sum_{\substack{J\leq S'\\|J|=j}}\frac{(n-i)!i!}{m^n}\\
&= \sum_{i=k+1}^np_i\binom{n}{i}^2\frac{(n-i)!i!}{m^n}\\
&= \frac{n!}{m^n}\sum_{i=k+1}^np_i\binom{n}{i}\\
&= \frac{n!}{m^n}\sum_{i=k+1}^nx^i(1-x)^{n-i}\binom{n}{i},
\end{align}
where first we use the fact that since our input and output are both collision-free, there are $\binom{n}{i}$ ways of choosing $I$ and $J$, then apply cancellation, and finally substitute the values of $p_i$. 
This gives our error bound for terms not in our approximation.

Next we shall consider the term $\Delta P_{\leq k}$. 
First we can note that $p_i' \geq p_i \forall i\leq k$, due to the normalisation of $p_i'$. 
Thus we can rewrite this term as
\begin{align}
\mean[\Delta P_{\leq k}] &= \sum_{i=0}^k(p_i'-p_i)\sum_{\substack{I\subseteq[n]\\|I|=i}}\mean\left[P_I[S']\right].
\end{align}
Using the same techniques for computing the Laplace expansion and calculating the expected value of permanents of Gaussian matrices, we can show that
\begin{align}
\mean[\Delta P_{\leq k}] &= \sum_{i=0}^k(p_i'-p_i)\binom{n}{i}^2\frac{(n-i)!i!}{m^n}\\
&= \frac{n!}{m^n}\sum_{i=0}^k(p_i'-p_i)\binom{n}{i}.
\end{align}

Next we expand $p_i'$:
\begin{align}
\sum_{i=0}^kp_i'\binom{n}{i} &= \frac{\sum_{i=0}^kx^i(1-x)^{n-i}\binom{n}{i}}{\sum_{i=0}^kx^i(1-x)^{n-i}\binom{n}{i}}\\
&= 1,
\end{align}
and use this expansion as well as the value of $p_i$ to calculate
\begin{align}
\mean[\Delta P_{\leq k}] &= \frac{n!}{m^n}\left(1 - \sum_{i=0}^kx^i(1-x)^{n-i}\binom{n}{i}\right)\\
&= \frac{n!}{m^n}\sum_{i=k+1}^nx^i(1-x)^{n-i}\binom{n}{i}.
\end{align}
Finally, we use these values and sum over all collision-free $S'$, of which there are $\binom{m}{n}\approx m^n/n!$ to bound our total variation distance for a Haar-random unitary as
\begin{equation}
\mean[\Delta P] \leq 2\sum_{i=k+1}^nx^i(1-x)^{n-i}\binom{n}{i}.
\end{equation}

It seems like there should be some room for improvement in this bound. 
In particular, the use of the triangle inequality suggests that there might be more precise approximations of the expected distance.

\section{Worst case error}
\label{sec:worst-case}

We shall now give an improved error bound, using a different technique. This bound will be an improvement in two ways: first by improving the error bound by a factor of two, and second by being a bound for the worst-case error of any linear-optical interferometer, rather than a bound for a Haar-random unitary interferometer. We do so by finding an upper bound for the trace distance between our ideal partially distinguishable state and the approximation that results from truncating at some $k$, the size of the largest indistinguishable set of particles.
As the trace distance is an upper bound for any POVM measurement, we know that this will provide an upper bound for the difference in distribution produced by any interferometer.

Denoting our truncated state
\begin{equation}
\rho_{\leq k,x} = \frac{\sum_{i=0}^kx^i(1-x)^{n-i}\sum_{\substack{I\subseteq[n]\\|I|=i}}\rho_I}{\sum_{i=0}^kx^i(1-x)^{n-i}\binom{n}{i}},
\end{equation}
where the denominator is a normalising factor, and similarly
\begin{equation}
\rho_{>k,x} = \frac{\sum_{i=k+1}^nx^i(1-x)^{n-i}\sum_{\substack{I\subseteq[n]\\|I|=i}}\rho_I}{\sum_{i=k+1}^nx^i(1-x)^{n-i}\binom{n}{i}}.
\end{equation}
We can now estimate the trace distance by rewriting $\rho_{n,x}$ as
\begin{align}
\rho_{n,x}
&=\sum_{i=0}^kx^i(1-x)^{n-i}\binom{n}{i} \rho_{\leq k,x} \\
&\quad+\sum_{i=k+1}^nx^i(1-x)^{n-i}\binom{n}{i} \rho_{>k,x} ,
\end{align}
which, by convexity, gives
\begin{equation}
\tdist(\rho_{n,x},\rho_{\leq k,x}) \leq \sum_{i=k+1}^nx^i(1-x)^{n-i}\binom{n}{i}\label{eqn:error-sum}.
\end{equation}

If we want this bound to be small, we can use techniques like those used for working out the tails of the binomial distribution. 
For example, it is known that for $nx<k<n$ that
\begin{equation}
\tdist(\rho_{n,x},\rho_{\leq k,x}) \leq \textrm{exp}\left(-nD\left(\frac{k}{n}||x\right)\right),\label{eqn:worst-case}
\end{equation}
where $D(k/n||x)$ is the relative entropy between coins with bias $k/n$ and $x$, respectively \cite{arratia1989}. 
Choosing a value of $k=n\alpha$ for $\alpha>x$ will give an error bound of $\textrm{exp}(-nD(\alpha||x))$. 
Thus, choosing such a value for $k$ would give an error that decreases as $n$ increases, albeit at the cost of needing to increase $k$ linearly with $n$. 
While this is still an exponential time algorithm, it would offer a polynomial speedup over the Clifford \& Clifford method for Boson Sampling with indistinguishable photons.

We can also note that the trace distance is only dependent on the initial states and not the measurement outcomes. 
As a result, this error bound also applies in the case where the output is not collision free.

\section{Incorporating loss}
\label{sec:loss}

We now consider how to adapt this simulation to Boson Sampling under uniform loss. We shall assume that each photon survives with probability $\eta$.

In \cite{oszmaniec2018,moylett2018}, it was shown that the initial state for Boson Sampling with a fixed number of lost photons can be represented in the first quantisation as the initial state
\begin{equation}
\frac{1}{\binom{n}{\ell}}\sum_{\substack{L\subseteq[n]\\|L|=\ell}}\frac{1}{\ell!}\sum_{\sigma,\sigma'\in\symm_n}\sigma\ket{s_L}\bra{s_L}\sigma'^\dagger,
\end{equation}
where $s_L$ is the state where photons in the subset $L$ of the original input photons have survived. 
In order to generalise this to uniform loss, we append $n-\ell$ "lost" photons in an additional spatial mode (single particle state $0$) which isn't affected by the interferometer:
\begin{align}
\left(\frac{1}{\binom{n}{\ell}}\sum_{\substack{L\subseteq[n]\\|L|=\ell}}\frac{1}{\ell!}\sum_{\sigma,\sigma'\in\symm_n}\sigma\ket{s}\bra{s}\sigma'^\dagger\right)\otimes(\ket{0}\bra{0})^{\otimes n-\ell}.
\end{align}
Note that in the same way that it doesn't matter which particles are traced out when initially applying the loss, it similarly doesn't matter which particles are replaced with the $\ket{0}\bra{0}$ state. 
Uniform loss matches that of choosing which subset of photons survive according to the binomial distribution \cite{oszmaniec2018,renema2018loss}. 
We can combine this model with the distinguishability model of Sec.\ \ref{sec:expansion}, giving
\begin{widetext}
\begin{align}
\rho_{n,\eta,x}&=\sum_{\ell=0}^n\eta^\ell(1-\eta)^{n-\ell}\sum_{\substack{L\subseteq[n]\\|L|=\ell}}\sum_{i=0}^\ell x^i(1-x)^{\ell-i}\sum_{\substack{I\subseteq L\\|I|=i}}\left(\frac{1}{\ell!}\sum_{\substack{\sigma,\sigma'\in\symm_n\\\sigma^{-1}(j)=\sigma'^{-1}(j)\forall j \notin I}}\sigma\ket{s}\bra{s}\sigma'^\dagger\right)\otimes(\ket{0}\bra{0})^{\otimes n-\ell}\\
&=\sum_{\ell=0}^n\eta^\ell(1-\eta)^{n-\ell}\sum_{\substack{L\subseteq[n]\\|L|=\ell}}\sum_{i=0}^\ell x^i(1-x)^{\ell-i}\sum_{\substack{I\subseteq L\\|I|=i}}\rho_{L,I}.
\end{align}
\end{widetext}

We can now see how our classical simulation for Boson Sampling under distinguishability can be adapted to accommodate loss as well. 
First, we choose a subset of photons $L$ to indicate the photons that were not lost. 
From this subset, we choose another subset of photons $I\subseteq L$ to indicate the indistinguishable photons, which are simulated via the Clifford \& Clifford algorithm. 
The photons in $L\setminus I$ are all distinguishable photons, and can be simulated classically as before. 
The classical complexity of this algorithm depends on the number of indistinguishable photons we choose. 
As in Section \ref{sec:expansion}, by truncating this to be some maximum size $k$ we can get an algorithm that runs in $O(k2^k+\poly(k,m))$ time.

To understand the precision of this algorithm, we first note that if $|L|\leq k$, then we can classically simulate any number of indistinguishable photons within our desired runtime. 
As a result, we only need to truncate when $|L|>k$, and only need to do so up to $|I|\leq k$. 

As with Sec.\ \ref{sec:average-case}, we start by considering the error bound for a random interferometer. We note that in cases where at most $k$ photons survive our approximation is exact, so these outcomes do not contribute to our total variation distance. 
For the remainder, we see that
\begin{equation}
\mean[\Delta P] = \sum_{\ell=k+1}^n\eta^\ell(1-\eta)^{n-\ell}\sum_{\substack{L\subseteq[n]\\|L|=\ell}}\mean[\Delta P_L],
\end{equation}
where $\Delta P_L$ denotes the error of our simulation with photons in input modes denoted by $L$. 
Using our bound in Sec.\ \ref{sec:average-case} as well as the rule of conditional binomial distributions (see below), we can bound this as
\begin{equation}
\mean[\Delta P] \leq 2\sum_{i=k+1}^n(\eta x)^i(1-\eta x)^{n-i}\binom{n}{i}.
\end{equation}

Again, an improvement over the use of the triangle inequality can lead to an improvement in this bound.

For the worst-case error, we construct analogous states to those in Sec.\ \ref{sec:worst-case}:
\begin{widetext}
\begin{equation}
\rho_{\leq k,\eta,x} = \frac{\sum_{\ell=0}^n\eta^\ell(1-\eta)^{n-\ell}\sum_{\substack{L\subseteq[n]\\|L|=\ell}}\sum_{i=0}^{\min(\ell,k)} x^i(1-x)^{\ell-i}\sum_{\substack{I\subseteq L\\|I|=i}}\rho_{L,I}}{\sum_{\ell=0}^n\eta^\ell(1-\eta)^{n-\ell}\binom{n}{\ell}\sum_{i=0}^{\min(\ell,k)} x^i(1-x)^{\ell-i}\binom{\ell}{i}},
\end{equation}

\begin{equation}
\rho_{>k,\eta,x} = \frac{\sum_{\ell=k+1}^n\eta^\ell(1-\eta)^{n-\ell}\sum_{\substack{L\subseteq[n]\\|L|=\ell}}\sum_{i=k+1}^\ell x^i(1-x)^{\ell-i}\sum_{\substack{I\subseteq L\\|I|=i}}\rho_{L,I}}{\sum_{\ell=k+1}^n\eta^\ell(1-\eta)^{n-\ell}\binom{n}{\ell}\sum_{i=k+1}^\ell x^i(1-x)^{\ell-i}\binom{\ell}{i}},
\end{equation}
\end{widetext}
and note that $\rho_{n,\eta,x}$ is a linear combination of these states. 
As a result, the worst-case error of this simulation, using the convex properties of the trace distance, can be bounded as
\begin{align}
\tdist(\rho_{n,\eta,x},\rho_{\leq k,\eta,x}) &\leq \sum_{\ell=k+1}^n\eta^\ell(1-\eta)^{n-\ell}\binom{n}{\ell}\nonumber\\
&\quad\times\sum_{i=k+1}^\ell x^i(1-x)^{\ell-i}\binom{\ell}{i}\\
&=\sum_{i=k+1}^n(\eta x)^i(1-\eta x)^{n-i}\binom{n}{i}\label{eqn:lossy-worst},
\end{align}
where in the second line we have used the rule of conditional binomial distributions. 
Using the same result as used for Eq.\ (\ref{eqn:worst-case}), we can bound the error to a constant value if $k=n\alpha$ for $\alpha>\eta x$ \cite{arratia1989}. 
This shows a relationship between distinguishability and loss similarly to, but not exactly the same, as the one found in \cite{renema2018loss}: the more distinguishable photons are, the more we can classically simulate photon loss, and vice versa. It is remarkable to see that these two algorithms have a different dependence on $x$ and $\eta$: while state truncation depends on $\eta x$, point truncation depends on $\eta x^2$. It is not immediately clear where this difference comes from, and we leave it as an open question.

\section{Empirical Errors}
\label{sec:empirical-errors}

A natural question at this point is how to assess the performance of this new approach over that of \cite{renema2018,renema2018loss}. 
It is not immediately clear how to find a fair comparison, as each approach has its own strengths and weaknesses. 
Truncating based on fixed points has the benefit of the error asymptotically tending towards a constant as $n$ increases, which means that $k$ can be chosen independently of $n$ and does not need to increase. 
But this comes at the cost of a potentially large, albeit polynomial, runtime of at least $O(n^{2k}k2^k(n-k)^4\log(n-k))$~\cite{renema2018,huber2008}. 
Truncating based on states, on the other hand, provides a significant improvement in runtime based on $k$, and is able to run in $O(2^k + \poly(k,m,n))$ time, but at the cost of $k$ increasing linearly with $n$ for constant error.

We therefore consider a variety of comparisons. 
In Sec.\ \ref{subsec:same-k}, we start by considering the highest value of $x$ and $\eta$ simulable by each approach when given the same values of $n$ and $k$. 
We then introduce the runtime for each algorithm in Sec.\ \ref{subsec:runtimes}, by comparing how fast they can simulate particular values of $x$ and $\eta$ for increasing $n$. 
Finally in Sec.\ \ref{subsec:same-runtime}, we compare the highest value of $x$ and $\eta$ simulable by both algorithms for a 90-photon Boson Sampling experiment, where $k$ is varying but under the condition that the algorithms have similar run times. The motivation for this is that $90$ photons has been suggested as strict upper bound for what is achievable using classical computation~\cite{dalzell2018}.

Before we go further, we make a few observations on the calculations of error bounds and runtimes used in this section. 
Rather than using the asymptotic error bounds for fixed point truncation, which assume $n\rightarrow\infty$, we have used bounds for finite $n$. 
This provides an improvement in the error of up to $1/\sqrt{e}$. 
For the runtime of fixed point truncation, we explicitly calculate $\sum_{i=0}^k\binom{n}{i}R(n,n-i)i2^i(n-i)^4\log(n-i)$, where $R(n,n-i)=\binom{n}{i}\lceil i!/e\rfloor$ is the number of permutations with $n-i$ fixed points, and we have assumed that computing every permanent of the $(n-i)\times(n-i)$ matrices with non-negative entries requires $O((n-i)^4\log(n-i))$ time
\footnote{Note that this computation could in the worst case take $O(n-i)^7\log^4(n-i)$ time, depending on the matrix sparsity~\cite{huber2008}.}. 
For Metropolised Independence Sampling, we choose the number of probabilities to approximate via state truncation as 100, which matches currently used ``burn in'' and ``thinning'' times\cite{neville2017}, though it is worth noting that this number could be different depending on the distribution of partially-distinguishable and lossy bosons. 
For the runtime of state truncation, we use $2k2^k + mk(k-1)/2 + m(n-k)$, where we have used the fact that the first term is approximately equivalent to computing two matrix permanents, the second term is the polynomial overhead of the Clifford \& Clifford approach, and the final term is the polynomial overhead of sampling the fully distinguishable photons \cite{clifford2017}. 
For these calculations, we have also assumed that $m=n^2$
\footnote{Note that $m\in O(n^2)$ is only sufficient to ensure that the probability of seeing collisions from a Haar-random interferometer are small \cite{arkhipov2011}. 
The classical hardness of Boson Sampling is also dependent on entries of $U$ being drawn independently with high probability. 
To ensure this, $m$ could be required to be as large as $n^6$ \cite{aaronson2011}. 
However, it is widely believed, and often referenced in Boson Sampling experiments, that $m=n^2$ should be sufficient \cite{aaronson2011,broome2013, spring2013, tillmann2013, crespi2013,carolan2015,wang2017,zhong2018}.}.

Finally, we note that in Secs.\ \ref{subsec:same-k} and \ref{subsec:same-runtime}, we only consider distinguishability and loss separately, by comparing the highest value of $x$ simulable in a lossless system, and the highest value of $\eta$ simulated in a full indistinguishable system. 
Ideally one would compare highest combinations of $\eta$ and $x$ which are classically simulable. 
However, doing so is complicated by the fact that both methods handle combinations of noise differently: point truncation handles them as the parameter $\eta x^2$, whereas state truncation handles them as $\eta x$. 
With this in mind, we plot values for distinguishability and loss separately, and note that, for the same performance, a reduction in one implies an increase in the other.

\subsection{Comparison at same level of truncation}
\label{subsec:same-k}

\begin{figure*}
\subfloat[\label{fig:max-x-samek}]{\includegraphics[width=0.45\linewidth]{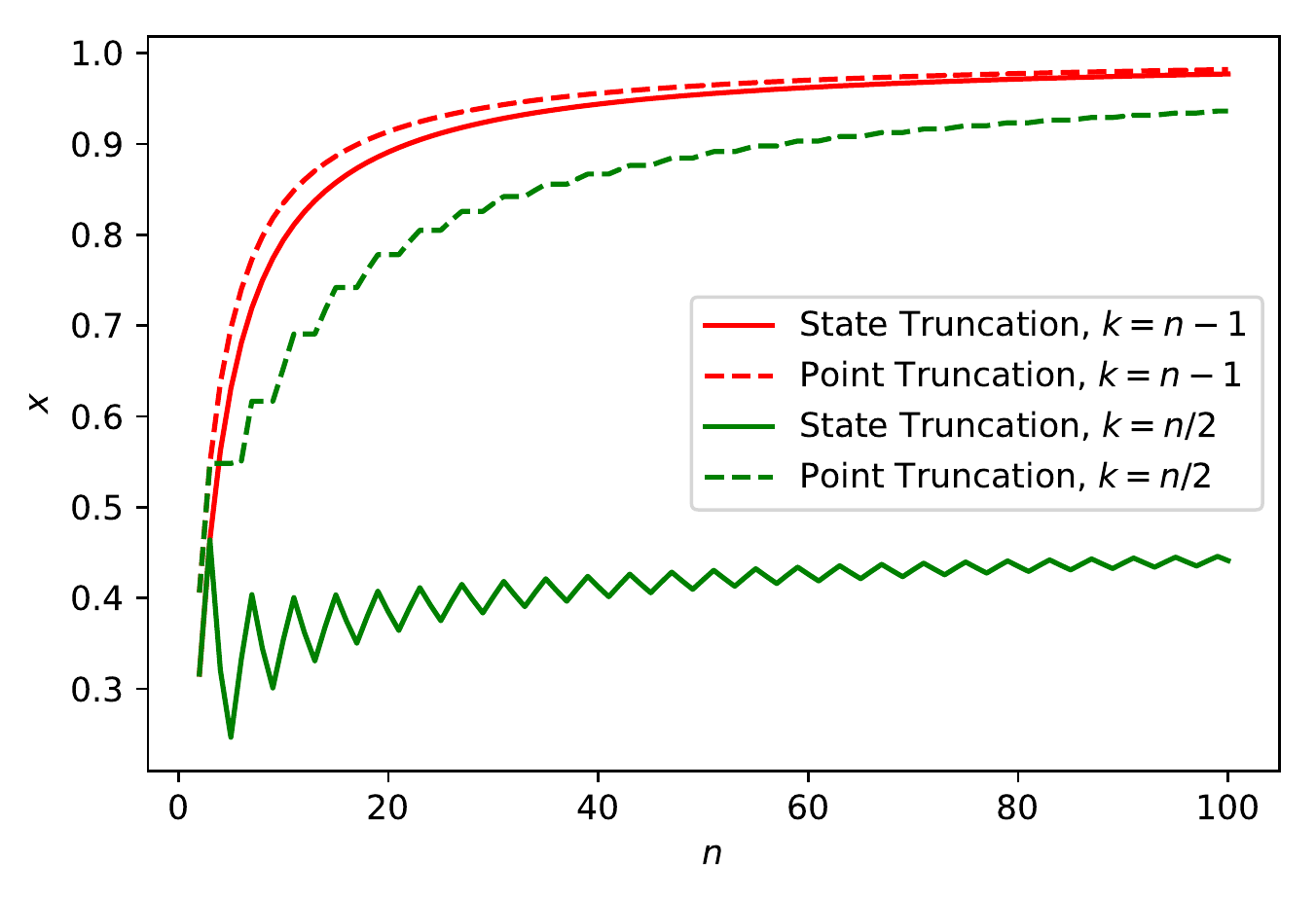}}
\hfill
\subfloat[\label{fig:max-eta-samek}]{\includegraphics[width=0.45\linewidth]{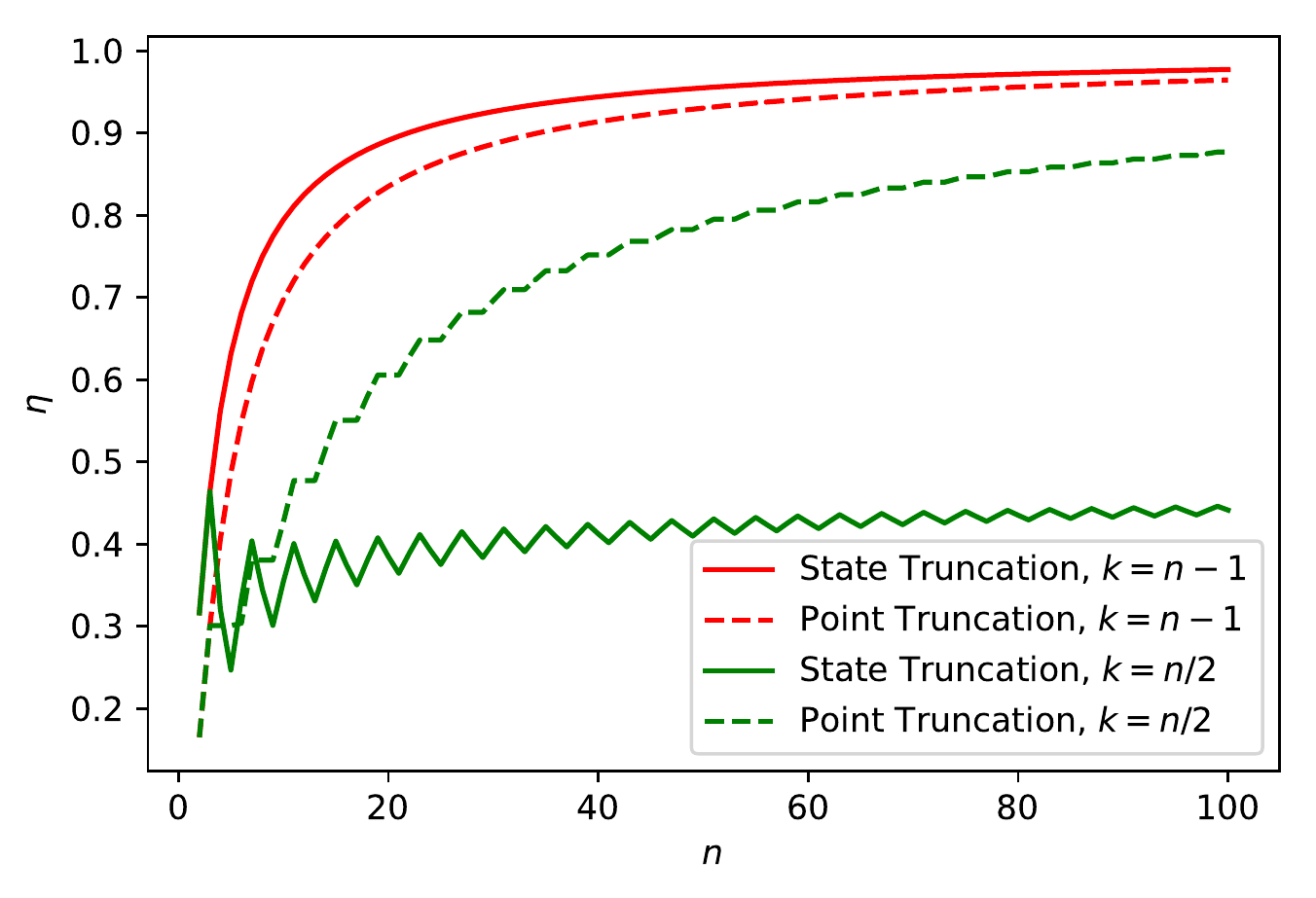}}
\caption{\label{fig:samek} 
Highest value of (\ref{fig:max-x-samek}) $x$ when $\eta=1$ and (\ref{fig:max-eta-samek}) of $\eta$ when $x=1$ simulable via state (solid) or point (dashed) truncation up to 10\% error ($\epsilon=0.1$). 
The number of photons, $n$, is varying, with $k$ chosen as either $k=n-1$ (red) or $k=n/2$ (green).
The oscillatory behaviour is due to rounding $k=n/2$ when $n$ is odd.}
\end{figure*}

We start by comparing the performance of the two algorithms when truncated at the same level $k$. 
This is of interest as in both approaches $k$ is considered to be a parameter defining the interference between photons. 
To do so, we consider the error bounds of classically simulating $n$-photon Boson Sampling for $n$ ranging between 2 and 100. 
The values chosen for $k$ depend on $n$: we consider $k=n-1$ as the upper limit of what the two algorithms can achieve without simulating the full distribution, and also $k=n/2$ as a more feasible, though still exponential time, value.

The result is plotted in FIG.\ \ref{fig:samek}, where in (\ref{fig:max-x-samek}) we show the highest value of $x$ simulable assuming no loss ($\eta=1$) and in (\ref{fig:max-eta-samek}) we show the highest value of $\eta$ simulable assuming the photons are fully indistinguishable ($x=1$). 
For all cases, we are considering simulations up to 10\% error.

There are a number of things we can note from FIG.\ \ref{fig:samek}. 
First is that when $k=n-1$, we can see that both algorithms tend to the same maximum values of distinguishability and loss. 
In the case of distinguishability, we can easily see why by considering the error bounds of both algorithms. 
One can see from Eq.\ (\ref{eqn:error-sum}) that state truncation will have a simple error bound in this case of $\epsilon \leq x^n$, meaning that for constant error the largest value of $x$ simulable is $x = \epsilon^{1/n}$. 
For point truncation, Eq.\ (\ref{eqn:renema-variance}) shows that the error is similarly bounded as $\epsilon \leq x^n/\sqrt{e}$, leading to a largest value of $x=(\epsilon\sqrt{e})^{1/n}$. 
Thus, although the highest value of $x$ simulable via point truncation is higher than that via state truncation, the difference will decrease in the limit of large $n$. 
Curiously we see the same effect as well in the case of loss, but now the highest value of $\eta$ simulable via state truncation is higher than that of point truncation. 
Again, this can be shown to hold theoretically: For state truncation the error scales as $\epsilon \leq \eta^n$ according to Eq.\ (\ref{eqn:lossy-worst}), corresponding to $\eta=\epsilon^{1/n}$; whereas for point truncation we see from Eq.\ (\ref{eqn:renema-variance}) and substituting $x=\sqrt{\eta}$ that the error scales as $\epsilon \leq \eta^{n/2}/\sqrt{e}$, meaning a maximum value of $\eta$ is $\eta=(e\epsilon^2)^{1/n}$. 
In the limit of large $n$ these differences will also tail off.

For $k=n/2$, we see that for both distinguishability and loss point truncation is more powerful than state truncation. 
Although this is harder to formally prove, there is intuition to see why this is the case. 
For state truncation, we know that for a small error to be achievable we need $k\geq n\eta x$, as this is the mean of the binomial distribution. 
Thus for $k=n/2$, we have that $\eta x \leq 1/2$, and in both cases we see the highest value of $x$ and $\eta$ tending to a value below $1/2$. 
For point truncation on the other hand, we know that the error tends to a constant value only dependent on $k$ and $\eta x^2$ in the limit of large $n$. 
As a result, it is unsurprising that for $k$ increasing linearly with $n$ the highest values of $x$ and $\eta$ will increase.

\subsection{Comparison of runtimes}
\label{subsec:runtimes}

\begin{figure}
\includegraphics[width=\linewidth]{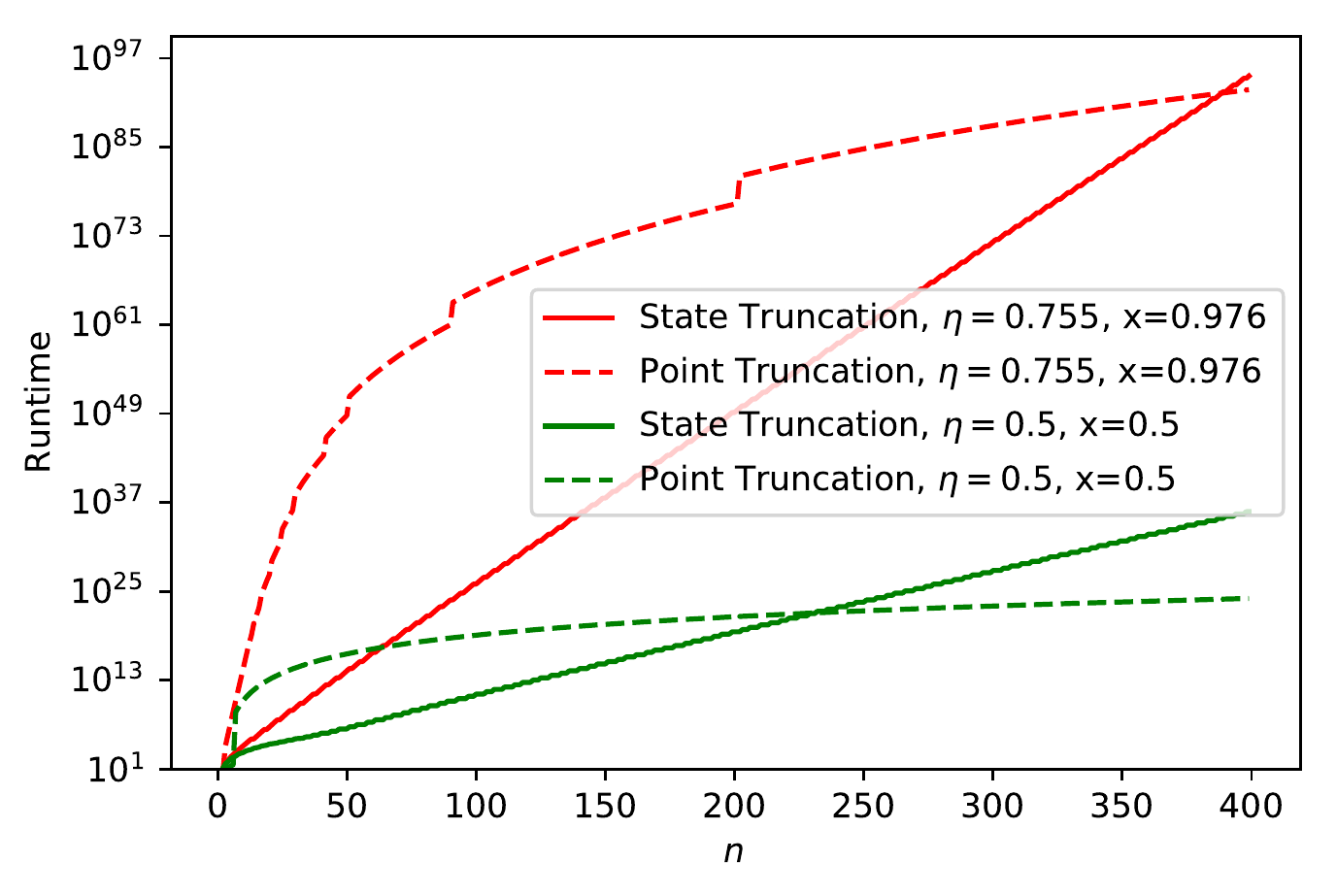}
\caption{\label{fig:runtime} Approximated runtime in terms of number of operations to simulate $n$-photon Boson Sampling with chosen values of $\eta$ and $x$ up to 10\% error ($\epsilon=0.1$) via state (solid) or point (dashed) truncation.}
\end{figure}

We next consider the runtime required to simulate $n$-photon Boson Sampling up to 10\% error via either method. 
The motivation for this comparison is that the runtime of the two algorithms at the same value of $k$ are significantly different. 
In particular, the runtime of state truncation is only dependent on $k$ and not $n$, whereas the runtime for point truncation depends on a scaling of approximately $O(n^{2k})$.

To understand how the runtimes scale, in FIG.\ \ref{fig:runtime} we plot the runtimes of classically simulating $n$-photon Boson Sampling experiments via the two approaches for fixed values of $\eta$ and $x$. 
The values of $k$ chosen for each algorithm are the smallest values for an error of at most 10\%. 
For choosing $\eta$ and $x$, we give two example cases. 
The first (FIG.\ \ref{fig:runtime}, red), where $\eta=0.755$ and $x=0.975$, is an example of a hypothetical best experiment we could build with current technology, with the most lossless sources (82\%) \cite{slussarenko2017}, interferometers (99\%) \cite{wang2018} and detectors (93\%) \cite{marsili2013}, and the highest level of photon indistinguishability (97.6\%) \cite{he2018}. 
The second case (FIG.\ \ref{fig:runtime}, green), where $\eta=x=0.5$, is an example of how the two algorithms perform in what would be considered a poor experiment for both distinguishability and loss. 
Actual Boson Sampling experiments are likely to fall between these two extremes.

In both cases, state truncation appears to outperform point truncation for near-term photon experiments, with point truncation eventually being able to perform faster for larger values of $n$. 
When $\eta=x=0.5$, point truncation performs better when $n$ is approximately larger than 230. 
In the case of $\eta=0.755, x=0.976$, point truncation only performs better when $n$ is approximately larger than $390$ photons. 
This gives an idea of the regions in which the polynomial runtime of point truncation can be better or worse than the exponential runtime of state truncation.

It is also worth noting that just because point truncation is faster than state truncation for large enough $n$ does not necessarily mean that either algorithm is efficient in these cases. 
When $\eta=x=0.5$, point truncation only becomes more efficient at instances where both algorithms already require the order of $10^{22}$ operations. 
And in the case where $\eta=0.755, x=0.976$, both algorithms have runtimes on the order of $10^{92}$ operations before point truncation outperforms state truncation.

\subsection{Comparison at same runtime}
\label{subsec:same-runtime}

\begin{figure*}
\subfloat[\label{fig:max-x-fixed-n-90}]{\includegraphics[width=0.45\linewidth]{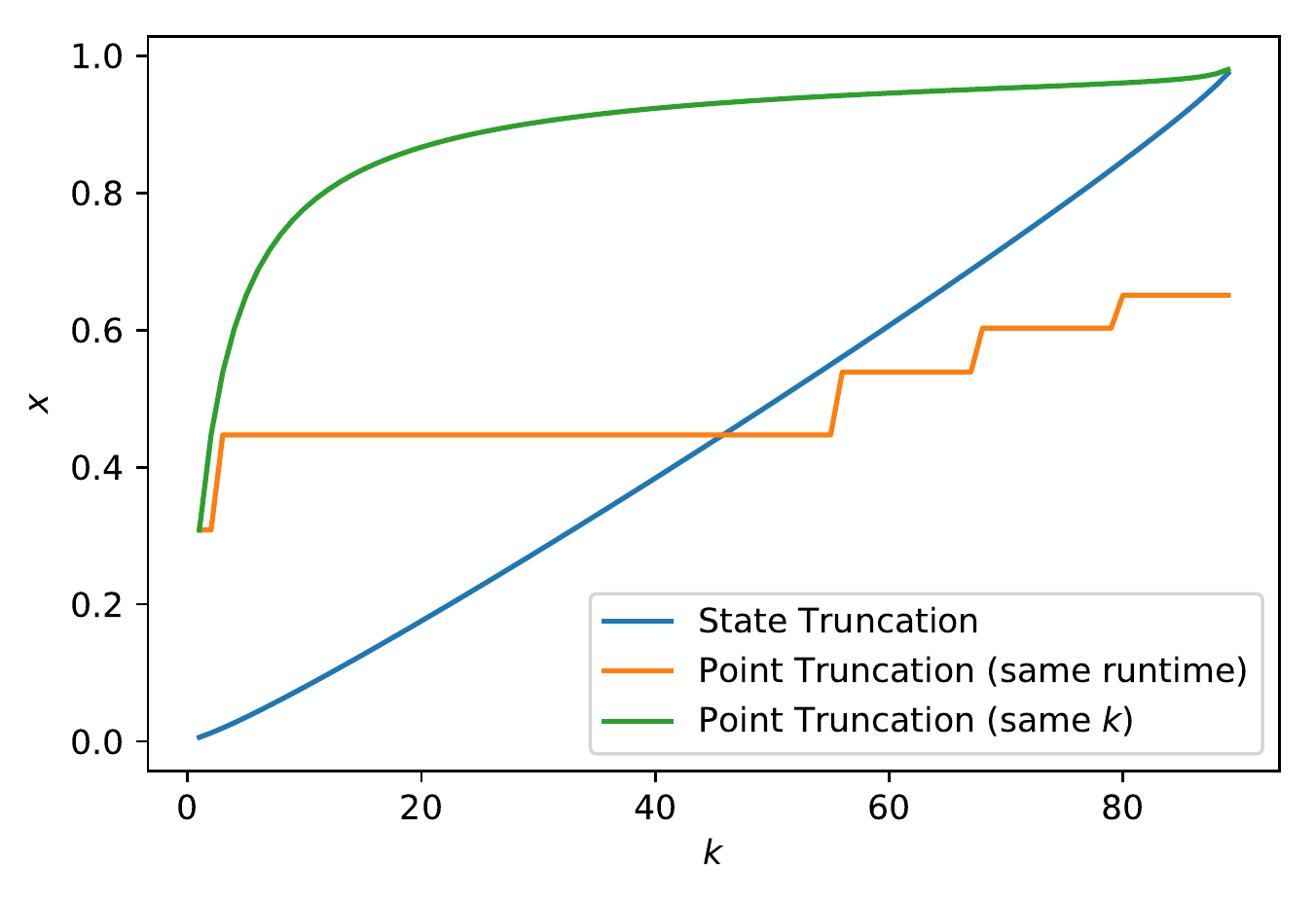}}
\hfill
\subfloat[\label{fig:max-eta-fixed-n-90}]{\includegraphics[width=0.45\linewidth]{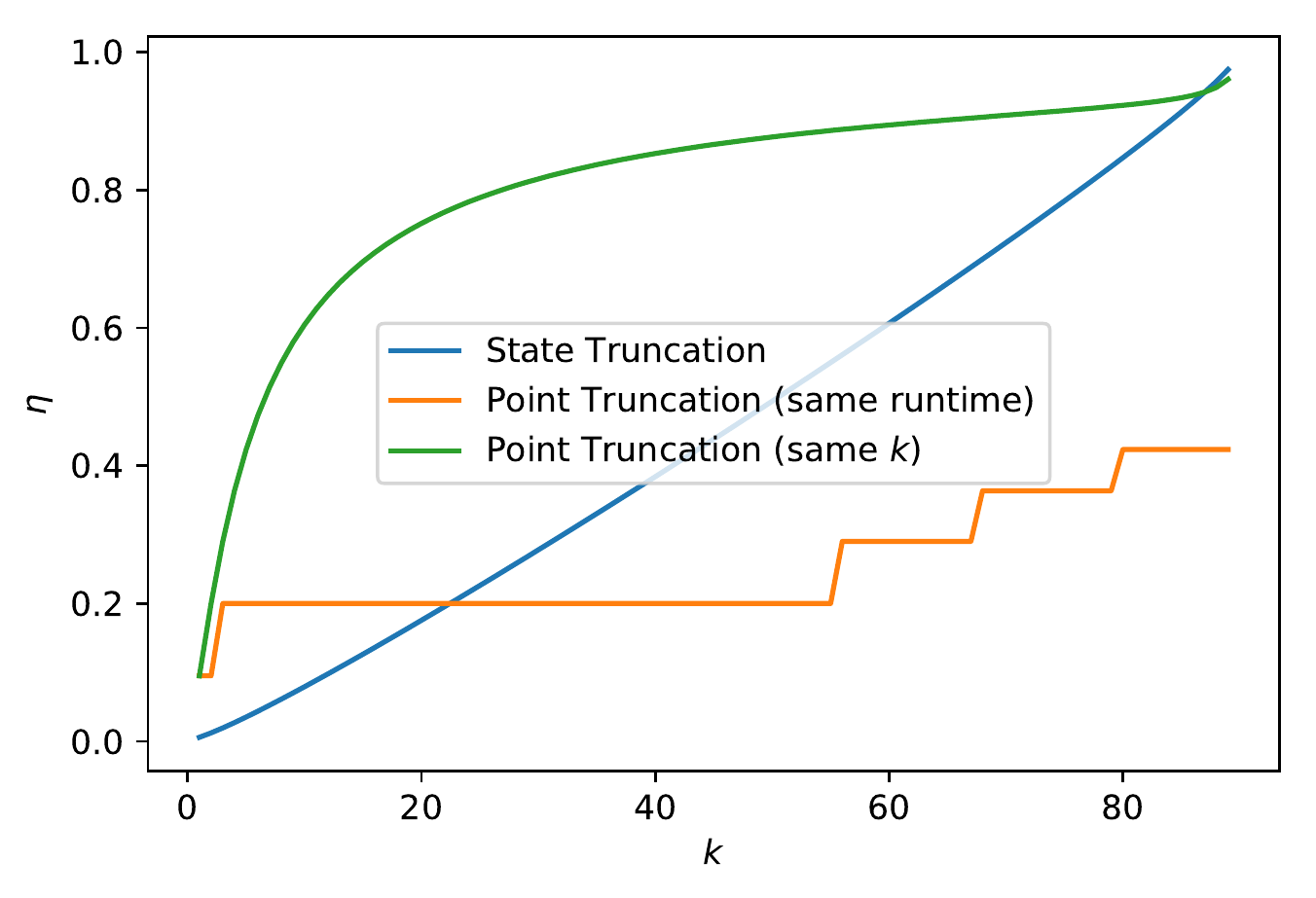}}
\caption{\label{fig:fixed-n-90} 
Highest value of (\ref{fig:max-x-fixed-n-90}) $x$ when $\eta=1$ and (\ref{fig:max-eta-fixed-n-90}) $\eta$ when $x=1$ simulable for 90-photon boson sampling at truncation level $k$ up to 10\% error ($\epsilon=0.1$). 
Blue line indicates highest values simulable via state truncation at level $k$, green lines indicate highest values simulable via point truncation at level $k$, orange lines indicate highest values simulable at point truncation at level $k'$ such that $k'$ is the smallest level of truncation such that the approximate runtime of point truncation at level $k'$ is longer than the runtime of state truncation at level $k$.}
\end{figure*}

Now we consider both truncation level and runtime, and compare the algorithms when restricted to comparable runtimes. 
To do this, we shall consider the challenge of simulating a 90-photon Boson Sampling experiment, and the largest values of distinguishability and loss that can be simulated at level $k$ with 10\% error. 
The motivation for this is that $90$ photons has been suggested as strict upper bound for what is achievable using classical computation~\cite{dalzell2018}. 

The results are shown in Fig.\ \ref{fig:fixed-n-90}, detailing for each algorithm the highest value of $x$ simulable when $\eta=1$ (\ref{fig:max-x-fixed-n-90}) and the highest value of $\eta$ simulable when $x=1$ (\ref{fig:max-eta-fixed-n-90}). 
In both figures, the blue line indicates state truncation at level $k$, and the green line indicates point truncation at level $k$. 
However, the runtime of state truncation at level $k$ and point truncation at level $k$ are likely to be drastically different. 
To take runtime into consideration as well, we consider the orange line which indicates point truncation at level $k'$, where $k'$ is the smallest integer such that the approximated runtime of point truncation at level $k'$ is longer than that of state truncation at level $k$. 
This allows us to compare the performance of the two algorithms when restricted to similar runtimes.

Considering distinguishability in FIG.\ \ref{fig:max-x-fixed-n-90}, we can note that point truncation with comparable runtime performs better up to $k\leq 45$, after which the methods are roughly comparable with state truncation performing marginally better, before becoming more dominant for $k\geq 60$. 
It has been suggested that boson sampling with 50 indistinguishable photons is roughly the limit of what can be classically simulated on a supercomputer \cite{neville2017, clifford2017, zhang2018}, so it appears that when considering distinguishability, the algorithms are roughly comparable in this case.

When considering loss in FIG. \ref{fig:max-eta-fixed-n-90}, we see a noticeable improvement for state truncation. 
Now point truncation under the same runtime only performs better up to $k\leq 22$, with state truncation performing considerably better for higher values of $k$. 
Boson Sampling with up to 30 indistinguishable photons is already known to be classically simulable on a standard laptop \cite{neville2017}, so this appears to offer a noticeable improvement even for fast classical simulations.

\section{Non-uniform loss}

We finish by briefly considering non-uniform loss, where each photon survives a lossy optical component with probability $\tau$. 
This model of loss has been considered before \cite{garciapatron2017,oszmaniec2018}, but without the incorporation of distinguishability. We can do this using the same methods as other non-uniform loss results, by extracting non-uniform losses into a layer of uniform losses followed by a lossy interferometer. 
The uniform loss layer means that each photon has probability $\eta=\tau^s$ of surviving, where $\tau$ is the loss of each optical component and $s$ is the smallest number of lossy optical components a photon interacts with. 
If we take the total number of lossy components to be $d$, the remaining lossy circuit can be modelled as an $(m+d)$-mode interferometer, with lost photons ending up in the additional $d$ modes. 
Thus we can achieve the same error as Eq.\ (\ref{eqn:lossy-worst}) in $O(k2^k+\poly(k,m,d))$ time. 
In typical schemes for linear interferometers, $d$ is at most polynomial in $m$ \cite{reck1994,clements2016}, so the overhead from these additional modes is small. 
We can bound the error to a constant if $k>nx\tau^s$. Taking the logarithm on both sides and rearranging for $s$, we find that this holds if

\begin{equation}
s>\frac{\log n-\log 1/x-\log k}{\log1/\tau}.
\end{equation}

This matches results in \cite{garciapatron2017,oszmaniec2018}, showing that boson sampling can be classically simulated if each photon encounters at least a logarithmic number of lossy components. 
It also shows how distinguishability can affect the simulability of lossy components in Boson Sampling: if our photons are more distinguishable, corresponding to a smaller value of $x$, then we can simulate shallower (i.e.\ less total loss) optical circuits.

\section{Conclusion}
\label{sec:conclusion}

In recent years significant improvements have been made in the ability of classical computers to simulate Boson Sampling under various imperfections. 
However, while it is of theoretical interest to demonstrate asymptotic improvements in classical simulation, the whole reason for proposals such as Boson Sampling is to offer speedups for near-term devices. 
Although our algorithm will not scale polynomially as the number of photons increases, we find that a substantial improvement over current classical algorithms can be achieved for the numbers of photons that experimentalists are currently aiming for. 
In doing so, we have effectively set a benchmark for what is required of a 50-90 photon Boson Sampling device.

There are a number of ways one could improve this classical simulation. 
In particular, the approach of Ref.~\cite{renema2018} for truncation when looking at near-term devices is dependent on Metropolised Independence Sampling. 
A direct adaptation of the Clifford \& Clifford algorithm to this approach would almost certainly offer an improvement over our algorithm. 
However, such an adaptation is non-trivial, due to the fact that the term in the expansion are not states, something that motivated our work here.

There are other open questions we would like to consider as well. The first would be to improve the average-case error bounds to match, if not improve over, our worst-case error bound. This would most likely involve an alternative to using the triangle inequality. The second would be to find a way of explaining the difference in dependence on $\eta$ and $x$ between point and state truncation, and ideally improving either algorithm in the process.

\subsection{Note Added}

During this work we were made aware of independent work by V. Shchesnovich, which also shows that the model of distinguishability considered by Renema \textit{et al.}\ corresponds to that of selecting indistinguishable photons via the binomial distribution \cite{shchesnovich2019}. 
This is derived using significantly different methods from those used in this manuscript, and does not consider classical simulation of distinguishability via the above method (though this has been anticipated~\cite{shchesnovich2019clifford}).

No underlying data was produced during this study.

\begin{acknowledgments}
A.E.M. was supported by the Bristol Quantum Engineering Centre for Doctoral Training, EPSRC grant EP/L015730/1.
R.G.P. Acknowledges the support of F.R.S.- FNRS and the Fondation Wiener Anspach.
J.J.R. acknowledges support from NWO Vici through Pepijn Pinkse.
We would like to thank Nicol\'{a}s Quesada for bringing the point about Gaussian Boson Sampling in Sec.\ \ref{sec:intro} to our attention.
\end{acknowledgments}

\bibliographystyle{apsrev4-1}
\bibliography{classical_sim}
\end{document}